\documentclass[useAMS,usenatbib,usegraphicx]{mn2e}
\usepackage{psfig}
\def\araa{ARA\&A}%
\def\apj{ApJ}%
\def\apjl{ApJ}%
\def\apjs{ApJS}%
\def\aap{A\&A}%
\def\aaps{A\&AS}%
\def\cjaa{Chinese J. Astron. Astrophys.}%
\def\mnras{MNRAS}%
\def\pasp{PASP}%
\def\ssr{Space~Sci.~Rev.}%
\def\nat{Nature}%
\title[Multiwavelength behaviour of the blazar OJ~248 from radio to $\gamma$-rays]{Multiwavelength behaviour of the blazar OJ~248 from radio to $\gamma$-rays.
}
\author[M.~I.~Carnerero et al.]
{M.~I.~Carnerero            $^{ 1,2,3}$,\thanks{E-mail:maribel@oato.inaf.it}
C.~M.~Raiteri               $^{ 1}$,
M.~Villata                  $^{ 1}$,
J.~A.~Acosta-Pulido         $^{ 2,3}$,
\newauthor
F.~D'Ammando                $^{ 4,5}$,
P.~S.~Smith                 $^{ 6}$,
V.~M.~Larionov              $^{ 7,8,9}$,
I.~Agudo                    $^{10,11}$,
\newauthor
M.~J.~Ar\'evalo             $^{ 2,3}$,
A.~A.~Arkharov              $^{ 8}$,
U.~Bach                     $^{12}$,
R.~Bachev                   $^{13}$,
E.~Ben\'itez                $^{14}$,
\newauthor
D.~A.~Blinov                $^{ 7,15}$,
V.~Bozhilov                 $^{16}$,
C.~S.~Buemi                 $^{17}$,
A.~Bueno~Bueno              $^{ 2,3}$,
\newauthor
D.~Carosati                 $^{18,19}$,
C.~Casadio                  $^{10}$,
W.~P.~Chen                  $^{20}$,
G.~Damljanovic              $^{21}$,
A.~Di Paola                 $^{22}$,
\newauthor
N.~V.~Efimova               $^{ 8}$,
Sh.~A.~Ehgamberdiev         $^{23}$,
M.~Giroletti                $^{ 5}$,
J.~L.~G\'omez               $^{10}$,
\newauthor
P.~A.~Gonz\'alez-Morales    $^{ 2,3}$,
A.~B.~Grinon-Marin          $^{ 2,3}$,
T.~S.~Grishina              $^{ 7}$,
M.~A.~Gurwell               $^{24}$,
\newauthor
D.~Hiriart                  $^{25}$,
H.~Y.~Hsiao                 $^{20}$,
S.~Ibryamov                 $^{13}$,
S.~G.~Jorstad               $^{ 7,26}$,
\newauthor
M.~Joshi                    $^{26}$,
E.~N.~Kopatskaya            $^{ 7}$,
O.~M.~Kurtanidze            $^{27,28,29,30}$,
S.~O.~Kurtanidze            $^{27}$,
\newauthor
A.~L\"ahteenm\"aki          $^{31,32}$,
E.~G.~Larionova             $^{ 7}$,
L.~V.~Larionova             $^{ 7}$,
C.~L\'azaro                 $^{ 2,3}$,
P.~Leto                     $^{17}$,
\newauthor
C.~S.~Lin                   $^{20}$,
H.~C.~Lin                   $^{20}$,
A.~I.~Manilla-Robles        $^{33}$,
A.~P.~Marscher              $^{26}$,
\newauthor
I.~M.~McHardy               $^{34}$,
Y.~Metodieva                $^{35}$,
D.~O.~Mirzaqulov            $^{23}$,
A.~A.~Mokrushina            $^{ 7,8}$,
\newauthor
S.~N.~Molina                $^{10}$,
D.~A.~Morozova              $^{ 7}$,
M.~G.~Nikolashvili          $^{27}$,
M.~Orienti                  $^{ 5}$,
\newauthor
E.~Ovcharov                 $^{16}$,
N.~Panwar                   $^{20}$,
A.~Pastor~Yabar             $^{ 2,3}$,
I.~Puerto~Gim\'enez         $^{ 2,3}$,
\newauthor
V.~Ramakrishnan             $^{31}$,
G.~M.~Richter               $^{36}$,
M.~Rossini                  $^{ 4}$,
L.~A.~Sigua                 $^{27}$,
\newauthor
A.~Strigachev               $^{13}$,
B.~Taylor                   $^{26,37}$,
M.~Tornikoski               $^{31}$,
C.~Trigilio                 $^{17}$,
\newauthor
Yu.~V.~Troitskaya           $^{ 7}$, 
I.~S.~Troitsky              $^{ 7}$,
G.~Umana                    $^{17}$,
A.~Valcheva                 $^{16}$,
S.~Velasco                  $^{ 2,3}$,
\newauthor
O.~Vince                    $^{21}$,
A.~E.~Wehrle                $^{38}$
and H.~Wiesemeyer           $^{12}$
}

\begin{document}


\pagerange{\pageref{firstpage}--\pageref{lastpage}} \pubyear{2015}

\maketitle

\label{firstpage}

\begin{abstract}
We present an analysis of the multiwavelength behaviour of the blazar OJ~248 at $z=0.939$ in the period 2006--2013. We use low-energy data (optical, near-infrared, and radio) obtained by 21 observatories participating in the GLAST-AGILE Support Program (GASP) of the Whole Earth Blazar Telescope (WEBT), as well as data from the {\it Swift} (optical--UV and X-rays) and {\it Fermi} ($\gamma$-rays) satellites, to study flux and spectral variability and correlations among emissions in different bands. We take into account the effect of absorption by the damped Lyman $\alpha$ intervening system at $z=0.525$. 
Two major outbursts were observed in 2006--2007 and in 2012--2013 at optical and near-IR wavelengths, while in the high-frequency radio light curves prominent radio outbursts are visible peaking at the end of 2010 and beginning of 2013, revealing a complex radio-optical correlation. Cross-correlation analysis suggests a delay of the optical variations after the $\gamma$-ray ones of about a month, which is a peculiar behaviour in blazars.
We also analyse optical polarimetric and spectroscopic data. The average polarization percentage $P$ is less than 3\%, but it reaches $\sim 19\%$ during the early stage of the 2012--2013 outburst. 
A vague correlation of $P$ with brightness is observed. There is no preferred electric vector polarisation angle and during the outburst the linear polarization vector shows wide rotations in both directions, suggesting a complex behaviour/structure of the jet and possible turbulence. The analysis of 140 optical spectra acquired at the Steward Observatory reveals a strong Mg II broad emission line with an essentially stable flux of $6.2 \times 10^{-15} \rm \, erg \, cm^{-2} \, s^{-1}$ and a full width at half-maximum of $2053 \rm \, km \, s^{-1}$.
\end{abstract}


\begin{figure*}
\centering
\includegraphics[width=14cm]{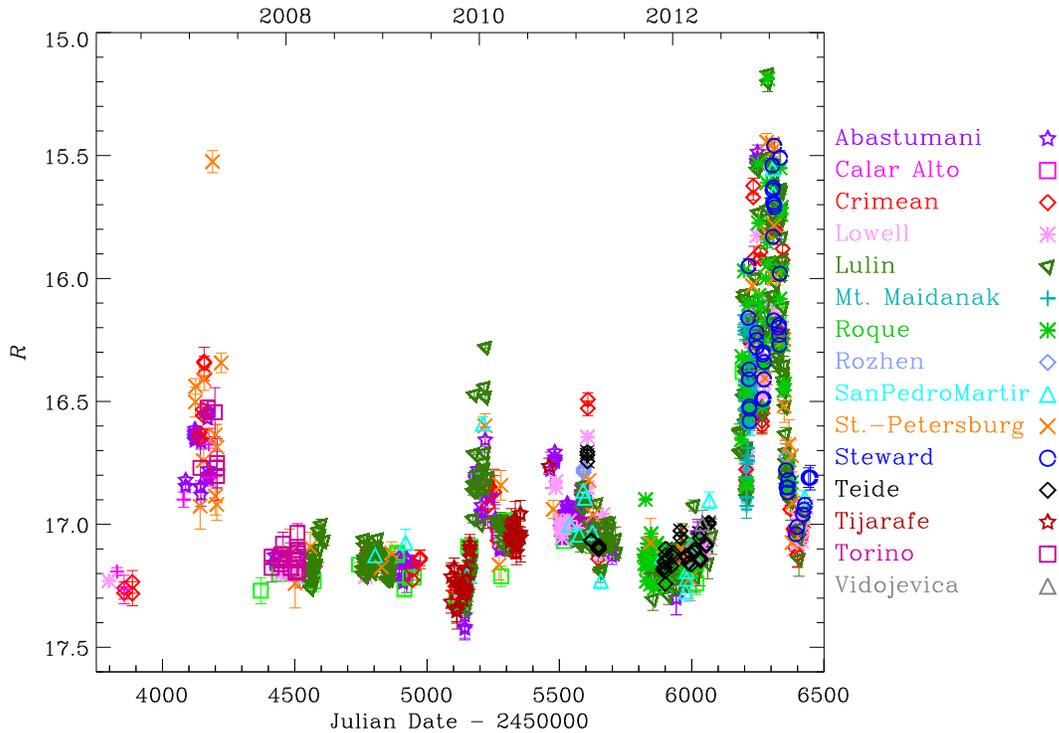}
  \caption{Light curve of OJ~248 in $R$ band. It includes 1211 points from 15 observatories, distinguished by different colours and symbols.}
  \label{optical}
\end{figure*}

\section{Introduction}

The emission of active galactic nuclei (AGNs) is explained by the existence of a supermassive black hole (SMBH) at the centre of the host galaxy, which converts gravitational energy of material located in the surroundings into electromagnetic energy. This material forms a disk and loses angular momentum due to the viscosity in the disk, finally falling onto the black hole. 
In general, the AGN spectra may show broad and narrow emission lines produced in regions close to the nucleus.
Sometimes they can also show lines from the host galaxy.
In radio-loud AGNs two plasma jets are ejected in direction perpendicular to the disc. 

Among the different types of radio-loud AGNs, the objects called ``blazars" (BL Lacs and flat spectrum radio quasars, FSRQs) are powerful emitters from radio wavelengths up to $\gamma$-ray energies. They present strong flux variability and high and variable polarization \citep[e.g.][]{smi96}.
The most accepted scenario to explain these features suggests that we are observing the emission from a jet of material accelerated to relativistic velocities in the vicinity of the SMBH, and oriented very close to our line of sight. Thus the jet radiation is Doppler boosted and dominates over the other emission components from the nucleus (disc, broad line region - BLR, narrow line region) or host galaxy. The origin of the low-frequency radiation (radio to UV or X-ray band) from the jet is attributed to synchrotron emission and the high-energy radiation (X- to $\gamma$-rays) to an inverse-Compton process by the same relativistic electrons producing the synchrotron photons.
After the launch of satellites for high-energy observations such as the {\it Astrorivelatore Gamma ad Immagini Leggero} \citep[AGILE;][]{tav09} and {\it Fermi} \citep{abd09_fermi,atw09}, the number of sources detected at $\gamma$-rays has increased significantly, allowing a more detailed investigation of the high-energy processes occurring in blazars.

Among blazars, in general BL Lacs are objects of lower luminosity with featureless spectra or very weak emission lines. In contrast, FSRQs have higher luminosities and stronger emission lines.
\begin{figure}
  \scalebox{1.0}{\includegraphics[width=8.5cm]{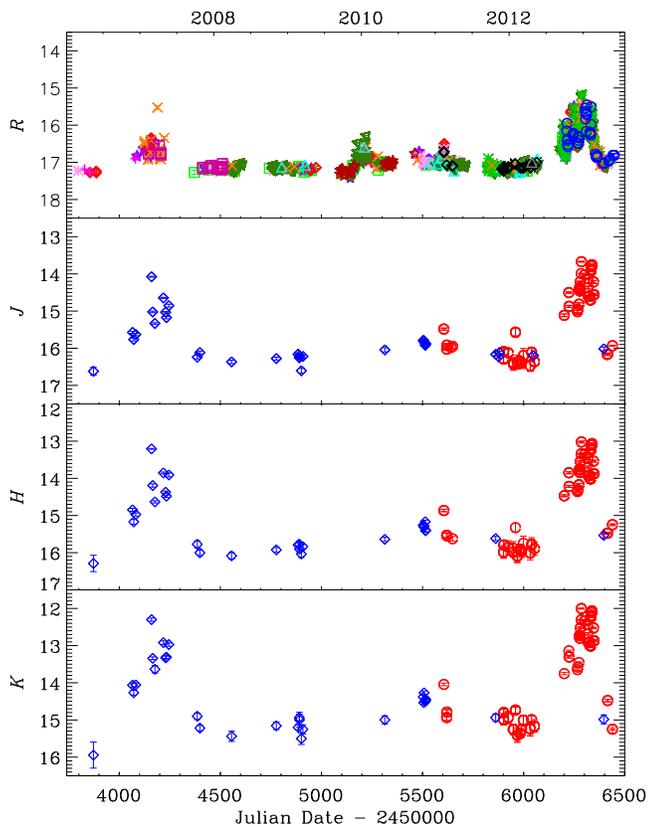}}
  \caption{Light curves of OJ~248 in optical ($R$ band) and near-IR ($J, H, K$ bands) in 2006--2013 built with GASP-WEBT data. In the near-IR light curves the blue diamonds are from the Campo Imperatore Observatory and the red points from the Teide Observatory.}
  \label{IR}
\end{figure}
In this paper, we present multifrequency observations of the FSRQ OJ~248 (0827+243) in 2006--2013 performed in the framework of a campaign led by the Whole Earth Blazar Telescope\footnote{http://www.oato.inaf.it/blazars/webt} (WEBT). The WEBT radio--optical observations are complemented by high-energy data from the {\it Swift} and {\it Fermi} satellites.

In the Roma BZCAT multi-frequency catalog of blazars\footnote{Edition 4.1.1, August 2012; http://www.asdc.asi.it/bzcat} \citep{mas09} OJ~248 appears with a redshift $z=0.939$ flagged as uncertain.
Mg II and Fe II absorption lines at $z=0.525$ were detected by \citet{ulr77} in the source optical spectrum. This intervening Damped Lyman $\alpha$ (DLA) system was subsequently studied by \citet{rao00}, who estimated a hydrogen column density $N_{\rm H} = (2.0 \pm 0.2) \times 10^{20} \rm \, cm^{-2}$. The DLA system is likely a disc galaxy with ongoing star formation \citep{ste02,rao03}. Although it does not affect the blazar photometry because of its faintness \citep{rao03}, its absorption of the source radiation must be taken into account.

OJ~248 was detected by the Energetic Gamma Ray Experiment Telescope (EGRET) instrument on board the {\it Compton Gamma Ray Observatory} (CGRO) with a variable flux.
In the third EGRET catalog of high-energy $\gamma$-ray sources \citep{har99} it appears with a flux $F(E>100 \rm \, MeV) = (24.9 \pm 3.9) \times 10^{-8} \, ph \, cm^{-2} \, s^{-1}$, single measurements ranging from 15.6 to $111.0 \times 10^{-8} \rm \, ph \, cm^{-2} \, s^{-1}$. It has $F(E>100 \rm \, MeV) = (5.3 \pm 0.5) \times 10^{-8} \, ph \, cm^{-2} \, s^{-1}$ and a spectral index $2.67 \pm 0.07$ in the Second Fermi-LAT catalog \citep[2FGL;][]{nol12}. 
In the Nineties the source was very active also in the optical band \citep[e.g.][]{vil97,rai98a}.

\section[]{Optical and near-IR photometry}
\label{onir}

Optical photometric observations were provided in the $R$ band by several observatories participating in this WEBT project, including the GASP-WEBT collaboration and the Steward Observatory program in support of the {\it Fermi} $\gamma$-ray telescope \citep{smi09}.
They are:
 Abastumani (Georgia, FSU),
 Calar Alto\footnote{Calar Alto data was acquired as part of the MAPCAT project: http://www.iaa.es/$\sim$iagudo/research/MAPCAT} (Spain),
 Crimean (Ukraine),
 Lowell (Perkins telescope, USA),
 Lulin (Taiwan),
 Mt.\ Maidanak (Uzbekistan),
 Roque de los Muchachos (Liverpool telescope, Spain),
 Rozhen (Bulgaria),
 San Pedro Martir (Mexico),
 St.\ Petersburg (Russia),
 Steward (USA),
 Teide (IAC80 telescope, Spain),
 Tijarafe (Spain),
 Torino (Italy),
 Vidojevica (Serbia).

The period of interest goes from 2006 March
up to 2013 July. 
We collected a total of 1356 data points, 1211 of which survived the light curve cleaning process, through which we discarded data with large errors as well as clear outliers. 


In the light curve of Fig.~\ref{optical} we can see two major flaring periods, in 2006--2007 and 2012--2013. The first flare is of similar brightness ($R=15.6$) as that observed in November 1995 by \citet{rai98a}, which was a historical maximum.  The second outburst appears more prominent and shows a stronger variability. Two minor events are visible in 2009--2010 and in early 2011.

The 2007 peak was very sharp and was characterized by a brightening of $\sim 1.3$ mag in 11 days and about 1.0 mag fading in 9 days.
A noticeable variability also characterises the 2012--2013 outburst, with variations of about 1 mag in less than 6 days.
This can be compared with variations of 1.43 mag in 16 days observed by \citet{rai98a}, of 1.16 mag in 63 days observed by \citet{vil97}, and of 1.05 mag in 58 days reported by \citet{fan04}.
But the source also exhibits short-term variability.
In particular, we found a couple of changes of $\sim 0.3$ mag in about 7.5 hours in late 2012. 
Intraday variability was previously reported by \citet{rai98a}, who observed a brightness decrease of 0.73 mag in 20 hours.

The GASP-WEBT near-IR data are collected in the $J, H, K$ bands at the Campo Imperatore (Italy) and Teide (TCS, Spain) observatories. Details on the data acquisition and reduction are given in \citet{rai14}. 
A comparison between the $R$-band and near-IR source behaviour is shown in Fig.\ \ref{IR}.
Although the near-IR light curves are less sampled than the optical one, we can recognize the same main features, in particular the two outbursts of 2006--2007 and 2012--2013.
The variation amplitude increases with wavelength, as it is usually observed in FSRQs \citep[e.g.][]{rai12}, suggesting the presence of a ``stable" blue emission component, likely
thermal radiation from the accretion disc.

\section{Optical polarimetry}

Blazars are known to show variable polarization in both polarized flux percentage ($P$) and electric vector polarization angle (EVPA) \citep[e.g.][]{smi96}. In particular, wide rotations of the linear polarisation vector have been detected in a number of cases \citep{mar08,mar10,lar13,sorcia14}, which have been interpreted as the effect of motion along spiral trajectories.

\begin{figure}
  \scalebox{1.0}{\includegraphics[width=9cm]{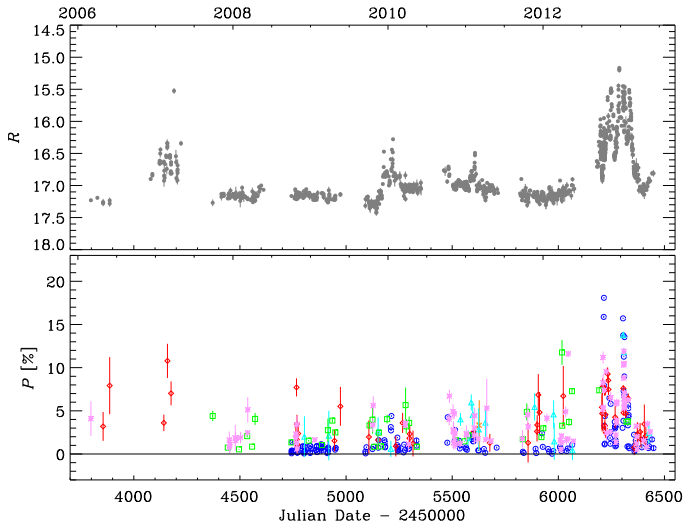}}
  \caption{Time evolution of the optical magnitude in $R$ band (top panel) and of the percentage of polarized flux (bottom panel). The data are from different observatories: Calar Alto (green squares), Crimean (red diamonds), Lowell (pink asterisks), San Pedro Martir (cyan triangles), Steward (blue circles), and St. Petersburg (orange crosses). 
}
  \label{pola}
\end{figure}

\begin{figure}
  \scalebox{1.0}{\includegraphics[width=8.5cm]{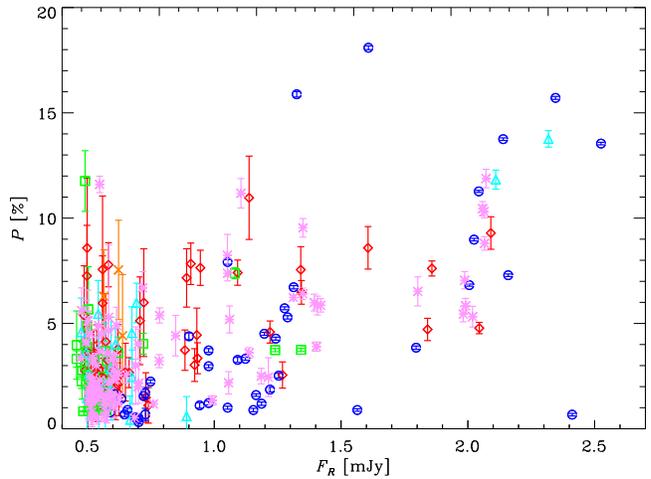}}
  \caption{Optical polarization percentage $P$ plotted against the de-absorbed flux density in the $R$ band. The data are from different observatories: Calar Alto (green), Crimean (red), Lowell (pink), San Pedro Martir (cyan), Steward (blue), and St. Petersburg (orange).}
  \label{polopt}
\end{figure}

Optical polarization data for this paper were provided by the Calar Alto, Crimean, Lowell, San Pedro Martir, St. Petersburg, and Steward observatories.
In Fig.\ \ref{pola} we show the time evolution of the polarization percentage $P$ compared with the $R$-band light curve.
For most of the time, the source showed low $P$ (average value of $\sim 3\%$), but during the brightening phase of the 2012--2013 outburst $P$ reached $\sim 19\%$, suggesting a correlation between $P$ and brightness typical of FSRQs \citep[e.g.][]{smi96,rai13}. 
Figure \ref{polopt} shows $P$ versus the de-absorbed flux density in the $R$ band (see Sect.\ 8). 
For any value of $F_R$ there is a large dispersion of $P$, but the highest values of $P$ ($>13\%$) are reached when the source is bright ($F_R > 1.3 \rm \, mJy$). The linear Pearson's correlation coefficient is 0.60, indicating a marginal correlation.

\begin{figure}
  \scalebox{1.0}{\includegraphics[width=8.5cm]{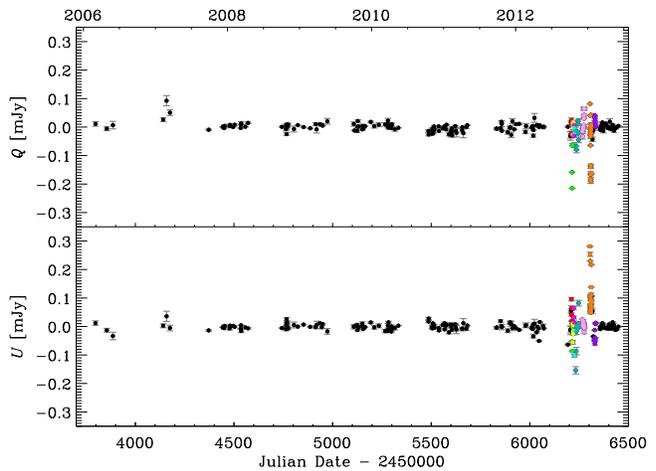}}
  \caption{The evolution of the $Q$ and $U$ Stokes'parameters as a function of time. Different colours during the 2012--2013 outburst highlight the data of the selected periods indicated in Fig.\ \ref{q_u}.}
  \label{stokes}
\end{figure}

\begin{figure}
  \scalebox{1.0}{\includegraphics[width=8.5cm]{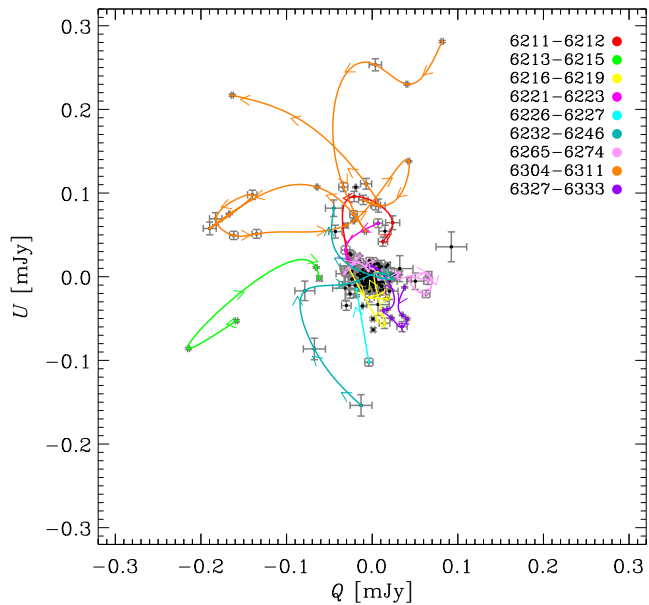}}
  \caption{$Q$ versus $U$ for all the data shown in Fig.\ \ref{stokes}. Coloured lines connect subsequent data belonging to the periods listed in the legend (JD-2450000). The direction is indicated by the arrows. }
  \label{q_u}
\end{figure}

\begin{figure}
  \scalebox{1.0}{\includegraphics[width=9cm]{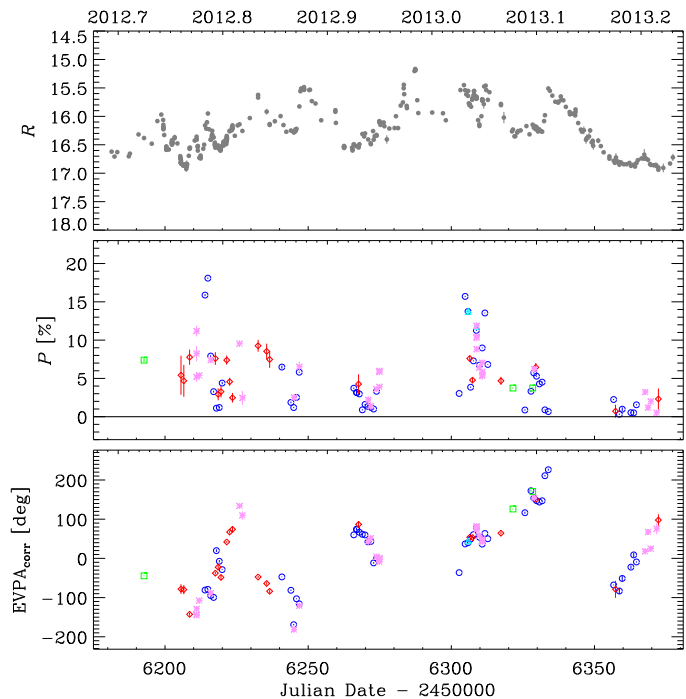}}
  \caption{Time evolution of the optical magnitude in $R$ band (top panel), of the percentage of polarized flux (middle panel) and of the EVPA ``corrected" for the $\pm \, n
 \pi$ ambiguity (bottom panel, see text for explanation) during the 2012--2013 outburst. Data are from different observatories: Calar Alto (green squares), Crimean (red diamonds), Lowell (pink asterisks), San Pedro Martir (cyan triangles), Steward (blue circles), and St. Petersburg (orange crosses). 
}
  \label{pola_zoom}
\end{figure}

In order to investigate whether possible rotations of the linear polarisation vector occurred, we first examine the behaviour of the $Q$ and $U$ Stokes' parameters.
Figure \ref{stokes} shows that OJ~248 spent most of the time with $Q$ and $U$ being close to zero.  With this being the case, even small variations in $Q$ and $U$ necessarily lead to large EVPA rotations that are difficult to accurately follow unless the observations are very dense. The problem is mitigated during the 2012--2013 outburst, when $Q$ and $U$ exhibit large variations. This appears clearer in the $Q$ versus $U$ plot in Fig.\ \ref{q_u}, where subsequent data belonging to short time periods with good sampling have been connected to show the time evolution during the 2012--2013 outburst.
With this in mind, in Fig.\ \ref{pola_zoom} we finally plot the EVPA as a function of time during the outburst.
The $\pm \, n \pi$ ambiguity was fixed by assuming that the most likely value is that minimizing the angle variation, i.e.\ 
we added/subtracted 180\degr\ when needed to minimize the difference between subsequent points separated by less than 5 days.
It seems that there is not a preferable EVPA value and that the linear polarisation vector underwent wide rotations in both directions and with different ranges of angles. All this suggests a complex behaviour/structure of the magnetic field in the jet as is expected e.g.\ from a turbulent plasma flowing at a relativistic speed down the jet \citep{mar14}.

\section{Optical spectroscopy}

Several previous studies found that there is no correlation between the jet activity and the behaviour of the broad emission lines in blazars \citep[see e.g.][]{cor00,rai07a}. Indeed, the BLR gas is likely ionized by the accretion disc radiation, whose variability is weaker and occurs on longer time scales than the variability of the beamed jet emission \citep[see e.g.][]{kas00}.
However, \citet{leo13} detected a flare-like variability of the Mg II emission line in the blazar 3C 454.3 during an outburst and claimed that the broad emission line fluctuations are linked to the non-thermal continuum emission from the jet. 
With this in mind, we analyse the spectroscopic behaviour of OJ~248 during our monitoring period.

Optical spectra were taken at the Steward Observatory of the University of Arizona for the ``Ground-based Observational Support of the {\it Fermi} Gamma-ray Space Telescope" program\footnote{http://james.as.arizona.edu/$\sim$psmith/Fermi/}.
Data for this program are taken at the 2.3~m Bok telescope and 1.54~m Kuiper telescope \citep{smi09}; 140 spectra of OJ~248 were acquired during the first five cycles of the {\it Fermi} mission, from October 2008 to June 2013. 

All spectra show a prominent Mg II $\lambda \lambda2796,2803$ broad emission line, which
was measured by fitting a Gaussian model with a single component, after
subtracting a linear fit to the continuum. We used  ad-hoc routines developed in IDL
and based on the MPFIT\footnote{http://www.physics.wisc.edu/$\sim$craigm/idl/fitting.html} libraries \citep{Markwardt09}.
The continuum region was selected from adjacent regions to the emission line, which are free of features.
The uncertainty of the measured flux was estimated from the average of the residuals obtained after the continuum fitting and was then used to determine the parameter confidence limits applying the routines in the MPFIT library.

\begin{figure}
  \scalebox{1.0}{\includegraphics[width=8.5cm]{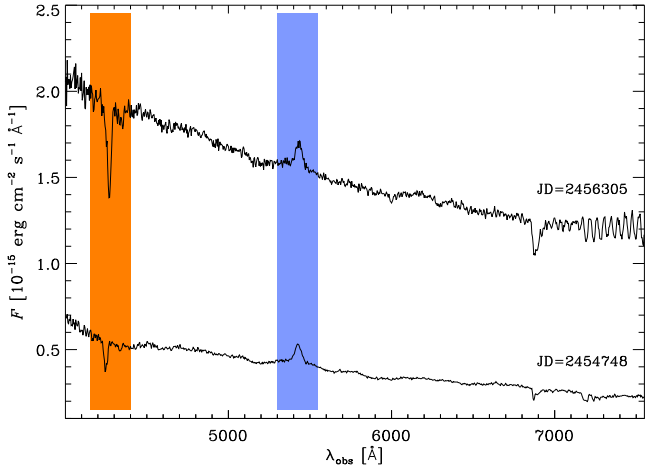}}
  \caption{Optical spectra of OJ~248 during different brightness states obtained at the Steward Observatory, showing the Mg~II emission line at $z=0.939$ (blue) and the Mg~II absorption line at $z=0.525$ (orange).}
  \label{espectro}
\end{figure}

\begin{figure}
\resizebox{\hsize}{!}{\includegraphics[angle=90]{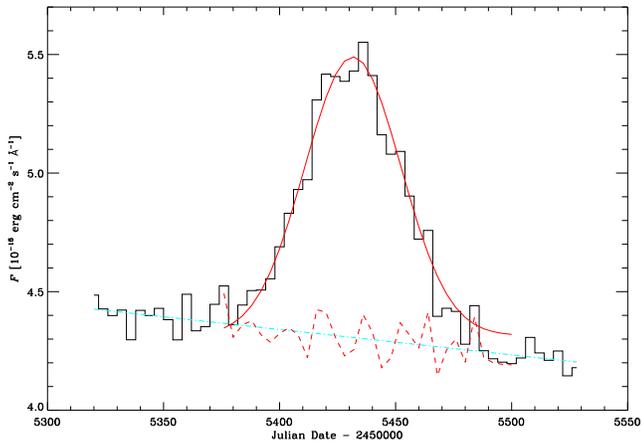}}  
  \caption{Gaussian fit (red solid line) to the Mg~II emission line of the spectrum taken on 2008 October 30 and shown in Fig.\ \ref{espectro}. The cyan line indicates the linear fit to the continuum; the red dashed line represents the residuals.}
  \label{fit}
\end{figure}

\begin{figure}
  \scalebox{1.0}{\includegraphics[width=8.5cm]{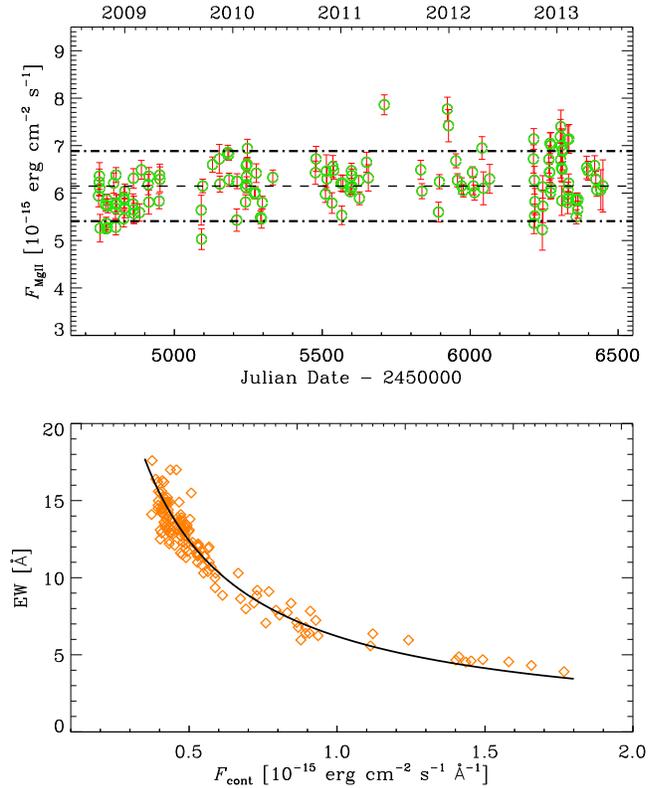}}
  \caption{Top: time evolution of the Mg II broad emission line flux; the dashed line represents the average value and the dotted-dashed line the standard deviation around the mean. Bottom: EW of Mg II versus the continuum flux density; the solid line displays the behaviour of the EW assuming a constant line flux equal to its average value.}
  \label{graf_espectro}
\end{figure}

In Fig.\ \ref{espectro} we show two of the spectra corresponding to different brightness states and in Fig.\ \ref{fit} the line and continuum fits as well as the residuals.
The results of the Mg II analysis are shown in Fig.\ \ref{graf_espectro}.
The line flux presents some dispersion around a mean value of $6.2 \times 10^{-15} \rm \, erg \, cm^{-2} \, s^{-1}$ with standard deviation of $0.5 \times 10^{-15} \rm \, erg \, cm^{-2} \, s^{-1}$. The possible presence of line variability can be checked e.g.\ by calculating the mean fractional variation 
$f={{\sqrt{\sigma^2-\delta^2}} / {<F>}}$,
where $\sigma$ is the standard deviation, $\delta$ the mean square uncertainty of the fluxes, and $<F>$ the average flux \citep{pet01}.
The result is $f=0.08$, which means that the line flux is basically stable, and this is true also during the 2012--2013 outburst period, when the continuum flux increased by a factor $\sim 6$. Moreover, no delayed line flux increase was detected also after the outburst. Hence, the enhanced jet activity responsible for the outburst does not affect the BLR.

We measured the line full width at half-maximum\footnote{Corrected for the instrumental broadening of the line.} (FWHM), from which one can derive the velocity of the gas clouds in the BLR. 
The average value is $v_{\rm FWHM} = 2053 \rm \, km \, s^{-1}$ with a standard deviation of $\sim 310 \rm \, km \, s^{-1}$.  The corresponding de-projected gas velocity, of course, depends on the geometry and orientation of the BLR \citep[see e.g.][]{dec11}.
In the blazar model, the BLR should be nearly face on, so measurements of FWHM are likely to be underestimated, since the measurement of the radial velocity will likely miss most of the orbital component. 

In the bottom panel of Fig.\ \ref{graf_espectro} we finally plot the equivalent width (EW) versus the continuum flux density. 
The EW decreases when the source brightens, which confirms that the jet is not the ionizing source of the BLR. 
We notice that, according to the classical definition \citep{sti91}, blazars with rest-frame EW less than 5 \AA\ are classified as BL Lacs; in the case of OJ~248, this happens when the observed EW goes below 5 \AA $\times (1+z) \sim 9.7$ \AA, which occurs when the source continuum flux density around the Mg II line\footnote{We estimated the continuum flux density in the spectral regions 5320--5360 \AA\ and 5500--5530 \AA\ and then took the mean value.} exceeds $\sim 0.6 \times 10^{-15} \rm \, erg \, cm^{-2} \, s^{-1} \, \AA^{-1}$. This underlines the limit of the classical distinction between BL Lacs and FSRQs based on the EW, which depends on the source brightness.

Fig.\ \ref{espectro} shows a strong absorption line at $\lambda=4270$ \AA\ due to Mg II absorption in the intervening DLA system at $z=0.525$ mentioned in the Introduction. We cannot perform a more detailed analysis of this line because the resolution of our spectra is about 20~\AA, implying a velocity resolution of $\sim 1400 \rm \, km \, s^{-1}$, while the FWHM of the Mg II absorption line is $\simeq 270 \rm \, km \, s^{-1}$ \citep{ste02}.

\section{Observations at radio and millimetre wavelengths}
\label{radiomm}

Radio and mm observations were performed with the Medicina (5, 8, and 22 GHz), Mets\"ahovi (37 GHz), Noto (43 GHz), IRAM (86 and 230 GHz), and Submillimeter Array (SMA, 230 and 345 GHz) telescopes.

A detailed description of the 43 GHz measurements performed with the Noto Radiotelescope can be found in \citet{let09}. For the Medicina observations, see e.g.\ \citet{bac07}.

The 37 GHz observations were made with the 13.7 m diameter Mets\"ahovi radio
telescope. The flux density scale is set by observations of DR 21. Sources NGC~7027,
3C~274 and 3C~84 are used as secondary calibrators. The error estimate in the flux 
density includes the contribution from the measurement root mean square and the uncertainty 
of the absolute calibration. A detailed description of the data reduction and analysis 
is given in \citet{ter98}.

IRAM 30~m Telescope data were acquired as part of the POLAMI (Polarimetric AGN Monitoring with the IRAM 30~m Telescope) and MAPI (Monitoring AGN with Polarimetry at the IRAM 30~m Telescope) programs. Data reduction was performed following the procedures described in \citet{agu06,agu10}.

Millimetre and submillimetre data were also obtained at the Submillimeter Array (SMA) near the summit of Mauna Kea (Hawaii). OJ~248 is included in an ongoing monitoring program at the SMA to determine the fluxes of compact extragalactic radio sources that can be used as calibrators at mm wavelengths \citep{gur07}. OJ~248 was also observed as part of a dedicated program to follow sources on the {\it Fermi} LAT Monitored Source List (PI: A. Wehrle) in 2009 and 2010. In the ongoing monitoring sessions, available potential calibrators are observed for 3 to 5 minutes, and the measured source signal strength calibrated against known standards, typically solar system objects (Titan, Uranus, Neptune, or Callisto). In addition, from time to time calibrator data obtained during regular science observations are also used to obtain flux density measurements. Data from this program are updated regularly and are available at the SMA website\footnote{http://sma1.sma.hawaii.edu/callist/callist.html}.

\begin{figure*}
\centering
\includegraphics[width=14cm]{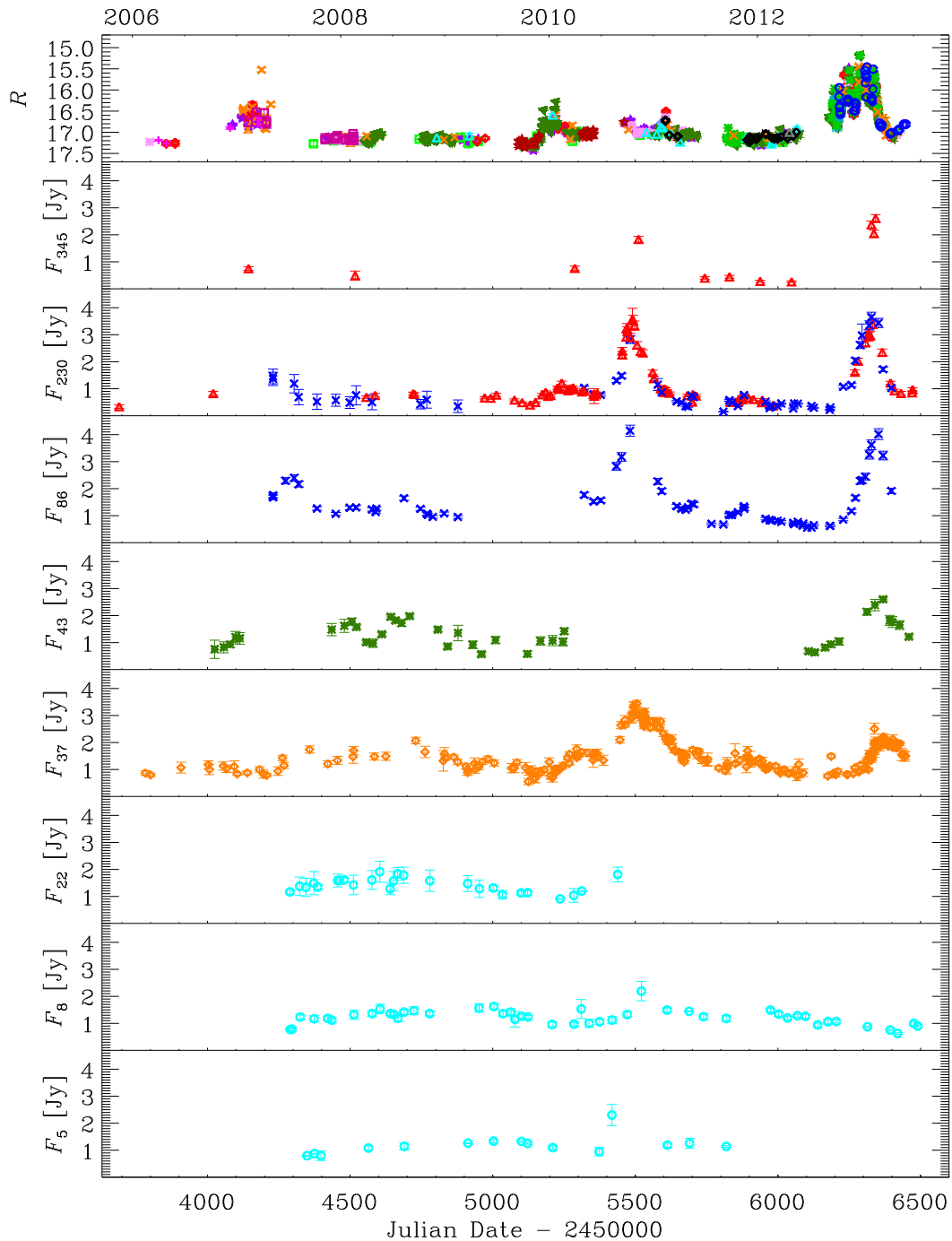}
  \caption{Light curves of OJ~248 at different frequencies in 2006--2013. From top to bottom: $R$-band optical magnitudes (see also Fig.\ \ref{optical}), 345 GHz data from SMA, 230 GHz data from SMA (red triangles) and IRAM (blue crosses), 86 GHz data from IRAM, 43 GHz data from Noto, 37 GHz data from Mets\"ahovi, 22, 8, and 5 GHz data from Medicina.}
  \label{radio}
\end{figure*}

\begin{figure}
\includegraphics[width=8.5cm]{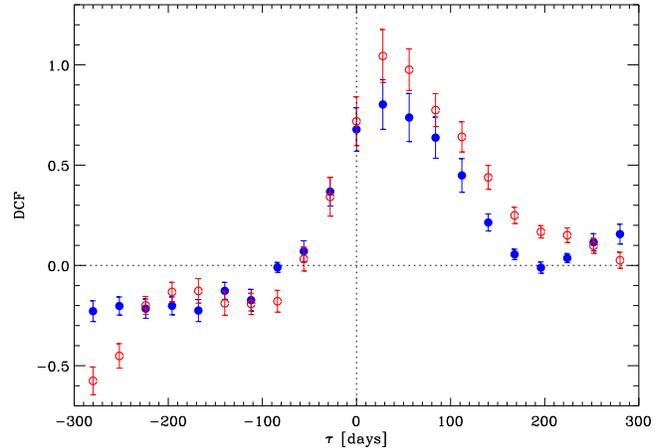}
  \caption{Discrete correlation functions between the $R$-band de-absorbed and 230 GHz flux densities (blue filled circles) and between the 230 and 37 light curves (red empty circles).}
\label{dcf_radio}
\end{figure}

As one can see in Fig.\ \ref{radio}, the mm data (230 and 86 GHz) show two prominent outbursts of the same strength peaking in late 2010 and early 2013.
Going to lower frequency (37 GHz), the second outburst becomes fainter than the first, and it completely disappears at 8 GHz. A comparison with the optical light curve suggests that the 2013 mm--radio outburst is the time delayed counterpart of the 2012--2013 optical event, while a possible correlation between the optical and radio variations in correspondence of the 2010--2011 mm--radio outburst is more difficult to establish. Indeed, there are no visible optical flares either contemporaneous or slightly preceding the lower-frequency outburst, but just one event 
about one year before, which however is more likely connected with the pre-outburst bumps visible at 230 and 37 GHz. In fact, there is a definite rise in the 230 GHz light curve that starts at essentially the same time as the late 2009--early 2010 optical event. 
Most likely the more or less prominent optical counterpart of the main mm--radio outburst remained unobserved due to the 2010 seasonal gap: indeed, some residual activity can be seen at the start of the subsequent observing season.
The 2006--2007 optical outburst might be correlated with a minor radio event observed a few months later at 86 (and 230) GHz.
We used the discrete correlation function \citep[DCF;][]{ede88,huf92} to analyse cross-correlations among light curves. Figure \ref{dcf_radio} shows the DCF between the optical de-absorbed (see Sect.\ \ref{asso}) and 230 GHz flux densities. The peak value is 0.8, which implies good correlation, and the time delay corresponding to the peak is 28 days, which would be the time lag of the mm variations after the optical ones. However, this result is dominated by the last outburst and the DCF run on the pre-outburst period gives no significant signal.

Flares can be produced by shocks propagating downstream the jet and/or by variations of the Doppler factor $\delta$, which depends on both the bulk Lorentz factor of the relativistic plasma, $\Gamma_{\rm b}$, and the viewing angle $\theta$, 
$\delta=[\Gamma_{\rm b} \, (1-\beta \, \cos\theta)]^{-1}$, where $\beta$ is the velocity in units of the light speed.
The different behaviour of OJ~248 in various epochs, i.e.\ different correlation between optical and radio variations, can be explained in terms of a misalignment of the region emitting the bulk of the optical radiation with respect to the zone emitting the bulk of radio photons \citep[see e.g.][]{vil07,vil09b,vil09a,rai11}.
The radiation coming from the jet region with a smaller viewing angle will in fact be more Doppler boosted. According to this interpretation, in 2010--2011 the radio emitting region was more aligned with the line of sight than the optical zone, while in 2012--2013 the most external jet regions, emitting the low-frequency radio photons, had a larger viewing angle, and the strong optical outburst was the effect of a viewing angle smaller than ever.

We finally notice that there is a time delay of the radio flux variations going toward longer wavelengths; in particular, by means of the DCF we could estimate the time lag between the 230 and 37 GHz flux changes  (see Fig.\ \ref{dcf_radio}). The peak of the DCF is strongly asymmetric, so that a better estimate of the delay in this case is given by the centroid of the distribution \citep{pet98}, which indicates a value of 40--50 days. Moreover, the amplitude of the flux variations decreases at lower frequencies (the ratio between the maximum and minimum flux density is $\sim 24$ at 230 GHz, 7.4 at 86 GHz, and 6.3 at 37 GHz), and the light curves become smoother with longer-lasting events. 
This is what we expect if the radio emission at longer wavelengths comes from more external (because of synchrotron self-absorption) and more extended regions of the jet.

An alternative picture to explain the radio--optical variability is in terms of a disturbance, e.g.\ shock wave, propagating downstream the jet where the delay from high radio frequencies towards low radio frequencies is naturally caused by opacity effects.
In case the optical and mm emitting regions are co-spatial, the reason why the mm peak is time delayed may be due to a rather high lower-energy cutoff to the electron energy distribution in the early stages of the outburst.

\section{{\it Swift} observations}

The {\it Swift} satellite \citep{geh04} carries three instruments that work simultaneously in different frequency ranges: the X-Ray Telescope (XRT), observing between 0.3 and 10 keV \citep{bur05}, the Ultraviolet-Optical Telescope (UVOT), between 170 and 600 nm \citep{rom05}, and the Burst Alert Telescope (BAT), between 14 and 195 keV \citep{bar05}.
OJ~248 was observed by {\it Swift} 86 times between 2008 January and 2013 May.
 
\subsection{UVOT observations}
\label{uvo}

The UVOT telescope can acquire data in optical ($v, b, u$) and UV ($w1, m2, w2$) bands \citep{poo08}.
The data reduction was performed with the {\sevensize \bf HEASoft} package version 6.13 and the Calibration Database (CALDB) 20130118 of the NASA's High Energy Astrophysics Science Archive Research Center\footnote{http://heasarc.gsfc.nasa.gov} (HEASARC). We extracted the source counts within a 5 arcsec radius aperture and the background counts from a nearby circular region with 15 arcsec radius. We summed multiple observations in the same filter with the {\tt uvotimsum} task and then processed them with {\tt uvotsource}.

\begin{figure}
  \scalebox{1.0}{\includegraphics[width=8.5cm]{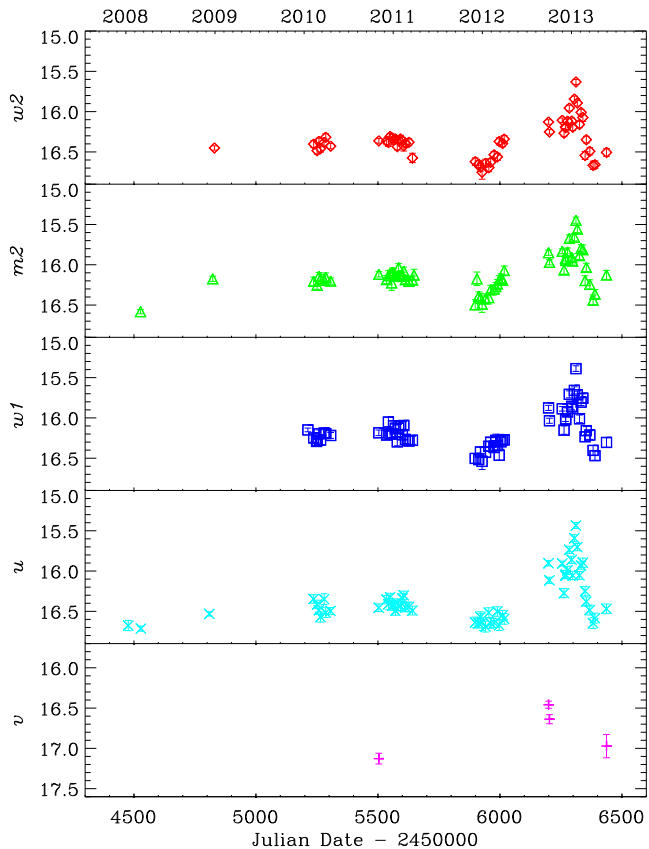}}
  \caption{Light curves of OJ~248 built with {\it Swift}-UVOT data in optical and UV.}
  \label{uvot}
\end{figure}

The resulting light curves are shown in Fig.~\ref{uvot}. No observations were available in the $b$ band. After 2010.0 the difference between the maximum and minimum magnitudes in the different bands is 1.12 in $w2$, 1.05 in $m2$, 1.15 in $w1$, 1.25 in $u$. It is 0.67 in $v$, but in this case we have only four points. We can see that the variability in general decreases when the frequency increases, extending the trend we noticed in Sect.\ \ref{onir} for the optical and near-IR light curves.

\subsection{XRT observations}

Reduction of the XRT data was performed with the {\sevensize \bf HEASoft package} version 6.13 with the calibration file 20130313. There are 83 observations in photon-counting (PC) mode. We ran the task {\tt xrtpipeline} with standard screening criteria.
For further analysis we only kept the 64 observations with more than 50 counts.
Source counts were extracted with the {\tt xselect} task from a circular region of 30 pixel radius centred on the source and the background counts were derived from a surrounding annular region of 50 and 70 pixel radii. No correction for pile-up was needed since the count rate is always lower than 0.5 counts/s. We performed spectral fits with the {\sevensize \bf Xspec} package in the 0.3--10 keV energy range, using the Cash statistics because of the low count number. 
We modelled the spectra with a power law with photoelectric absorption, adopting a hydrogen atomic column density $N_{\rm H}=4.6 \times 10^{20} \rm \, cm^{-2}$ obtained by summing the Galactic value $2.6 \times 10^{20} \rm \, cm^{-2}$ \citep{kal05} to that of the intervening DLA system at $z=0.525$ \citep{rao00}.

\begin{figure}
\resizebox{\hsize}{!}{\includegraphics[angle=-90]{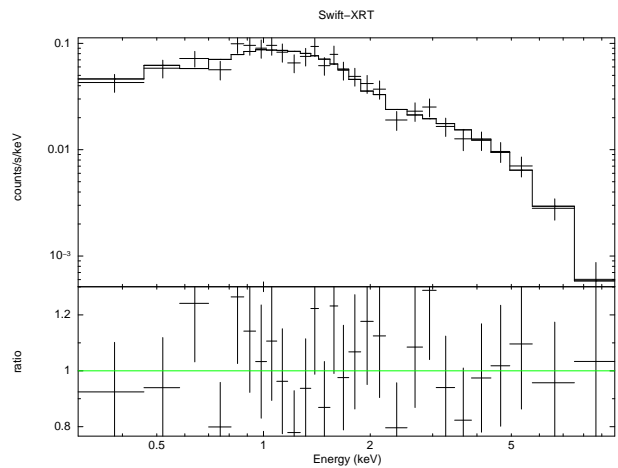}}  
  \caption{The XRT spectrum of OJ~248 on 2012 October 2. The best fit was obtained with a power law with fixed absorption given by the sum of the Galactic and intervening DLA $N_{\rm H}$ values. The bottom panel shows the ratio of the data to the folded model.}
\label{xrt}
\end{figure}

The X-ray spectrum acquired on 2012 October 2 is shown in Fig.~\ref{xrt} as an example. It was best-fitted with a power law with photon index $\Gamma=1.49 \pm 0.10$.

\begin{figure}
\includegraphics[width=8.5cm]{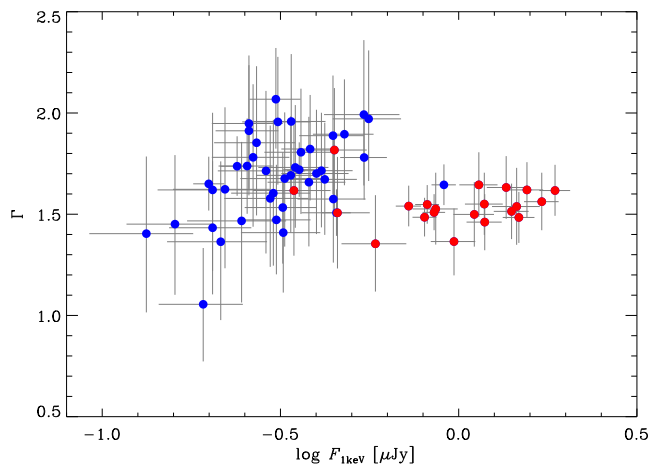}
  \caption{The X-ray photon index $\Gamma$ as  a function of the unabsorbed flux density at 1 keV. Only data with error less than 30\% of the flux are shown. Red points correspond to data acquired after $\rm JD=2456100$, i.e.\ during the 2012--2013 outburst.}
\label{xrtgamma}
\end{figure}

In Fig.~\ref{xrtgamma} we plotted $\Gamma$ as a function of the flux density at 1 keV for the 58 observations with error less than 30\% of the flux. The values of $\Gamma$ range between 1.06 and 2.07, with an average value of 1.65 and no significant trend of $\Gamma$ with flux, in contrast to the harder-when-brighter trend sometimes found in FSRQs \citep[e.g.][]{ver10}.
The smaller dispersion of the data points corresponding to the 2012--2013 outburst (standard deviation $\sigma=0.09$) with respect to the pre-outburst data ($\sigma=0.24$) is likely due to their higher precision because of the larger count number. In both cases the standard deviation is less than the average error (0.16 and 0.29, respectively), indicating that the data are consistent with a constant value. 

Notice that 
\citet{jor04} analysed Chandra data using $N_{\rm H}= 3.62 \times 10^{20} \rm \, cm^{-2}$, i.e.\ the Galactic value according to \citet{dic90}, and that
\citet{str13} found a value of $N_{\rm H}= (7\pm 2) \times 10^{20} \rm \, cm^{-2}$ when analysing XRT data with an absorbed power law with freely varying $N_{\rm H}$.

The resulting X-ray light curve (flux densities at 1 keV) is shown in Fig.\ \ref{multi}, where it is compared to the source behaviour at other frequencies.
 
\section{{\it Fermi} observations}

\begin{table*}
 \centering
  \caption{Results of the spectral analysis of the {\it Fermi}-LAT data of OJ~248 in the 0.1--100 GeV energy range. The reference energy $E_{\rm 0}$ was fixed to 392.1 MeV. The fitted model was a power law with photon index $\Gamma$.}
   \label{fermitab}
  \begin{tabular}{@{}crrcr@{}}
  \hline
Date &  $N_{\rm pred}$  &TS & $\Gamma$ & $F_{\rm 0.1-100 GeV}$ \\
     &                  &   &          & $[10^{-8} \rm \, ph \, cm^{-2} \, s^{-1}]$ \\
 \hline
2008 Aug 04 - 2009 Mar 05 &539 & 125 & $2.77 \pm 0.12$ & $6.2 \pm 0.8$\\
2009 Mar 05 - 2009 Oct 15 &465 & 131 & $2.55 \pm 0.11$& $5.2 \pm 0.8$\\
2009 Oct 15 - 2010 May 27 &406 & 104 & $2.60 \pm 0.12$& $5.3 \pm 0.8$\\
2010 May 27 - 2011 Jan 05 &466 & 100 & $2.83 \pm 0.12$& $6.6 \pm 0.9$\\
2011 Jan 06 - 2011 Aug 18 &439 & 97 & $2.81 \pm 0.13$& $6.4 \pm 1.0$\\
2011 Aug 18 - 2012 Feb 02 &283 & 43 & $3.13 \pm 0.23$& $5.5 \pm 1.0$\\
2012 Feb 02 - 2012 Jul 19 &392 & 107 & $2.66 \pm 0.13$& $6.9 \pm 1.0$\\
2012 Jul 19 - 2013 Jan 03 &1975 & 2106 & $2.38 \pm 0.03$& $31.5 \pm 1.3$\\
2013 Jan 03 - 2013 Jun 20 &730 & 401 & $2.48 \pm 0.07$& $12.6 \pm 1.1$\\
2013 Jun 20 - 2013 Nov 08 &313 & 103 & $2.56 \pm 0.13$& $6.5 \pm 1.1$\\
\hline
\end{tabular}
\end{table*}

The {\it  Fermi} satellite  was launched on 2008 June 11. Its aim is to perform a daily mapping of the $\gamma$-ray sources in the Universe. 
The primary instrument of {\it Fermi} is the Large Area Telescope \citep[LAT;][]{atw09}.
The energy range covered is approximately from 20 MeV to more than 300 GeV. The field of view of the LAT covers about 20\% of the sky, and maps all the sky every three hours.

The data in this paper were collected from 2008 August 4 ($\rm JD=2454683$) to 2013 November 8 ($\rm JD=2456605$). We performed the analysis with the {\sevensize \bf SCIENCETOOLS} software package version v9r32p5. The data were extracted within a Region of Interest (ROI) of 10\degr\ radius and a maximum zenith angle of 100\degr\ to reduce contamination from the Earth limb $\gamma$-rays, which are produced by cosmic rays interacting with the upper atmosphere. Only events belonging to the `Source' class were used. The time intervals when the rocking angle of the LAT was greater than 52\degr\ were rejected. For the spectral analysis we used the science tool {\tt gtlike} with the response function {\tt P7REP\_SOURCE\_V15}. 
Isotropic (iso\_source\_v05.txt) and Galactic diffuse emission
(gll\_iem\_v05.fit) components were used to model the
background\footnote{http://fermi.gsfc.nasa.gov/ssc/data/access/lat/\\BackgroundModels.html}.

We evaluated the significance of the $\gamma$-ray signal from the sources within the ROI by means of the Test Statistics  $\rm TS = 2 \, (\log L_1 - \log L_0)$, where $L_1$ and $L_0$ are the likelihood of the data given the model with or without the source, respectively \citep{mat96}. As was done in the 2FGL catalog \citep{nol12}, for the spectral modelling of OJ~248 we adopted a power law, $N(E)= N_{\rm 0} \, (E/E_{\rm 0})^{-\Gamma}$, where $E_{0}=392.1$ MeV is the reference energy between 0.1 and 100 GeV. We first ran {\tt gtlike} with the {\tt DRMNFB} optimizer, including all point sources of the catalog within 15\degr\ from our target, and using power-law fits to model the spectra of these sources. We then ran {\tt gtlike} a second time with {\tt NEWMINUIT} as optimizer, after selecting the sources with $\rm TS > 10$ and the predicted number of counts $N_{\rm pred}>3$.

\begin{figure*}
\centering
\includegraphics[width=14cm]{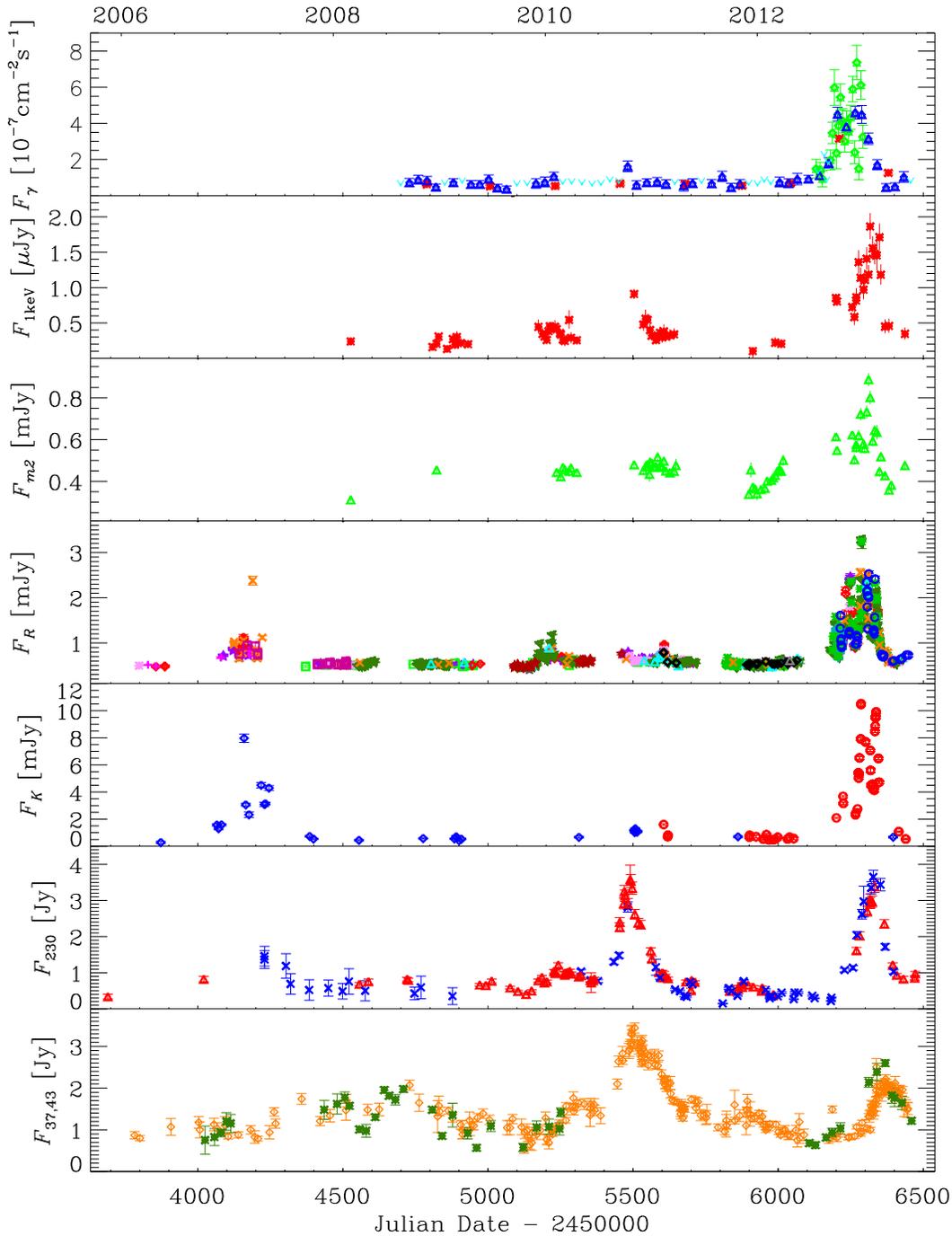}
  \caption{Light curves of OJ~248 at different frequencies in 2006--2013.
From top to bottom: the {\it Fermi}-LAT 0.1--100 GeV fluxes ($10^{-8} \rm \, ph \, cm^{-2} \, s^{-1}$) derived with different time bins (red symbols refer to data binned over roughly six-month time intervals, blue ones to monthly binned data, and green symbols to weekly-binned data in the outburst period; cyan arrows indicate upper limits); the 1-keV {\it Swift}-XRT flux densities ($\mu$Jy); the {\it Swift}-UVOT $m2$ flux densities (mJy); the GASP-WEBT $R$-band flux densities (mJy); the GASP-WEBT $K$-band flux densities (mJy); the 230 GHz flux densities (Jy); the 37 GHz (orange points) and 43 GHz (green points) flux densities (Jy). The X-ray, UV, optical and near-IR light curves were corrected by both Galactic and DLA absorption.}
\label{multi}
\end{figure*}

The results of the analysis are reported in Table \ref{fermitab} for time bins of about six months. 
The average photon index is 2.68, and its standard deviation is 0.22.
This value is essentially the same as that reported in the 2FGL catalog ($2.67 \pm 0.07$),
while an analysis over the whole period considered in this paper yields $\Gamma= 2.56 \pm 0.03$ 
and a flux of $(8.6 \pm 0.3) \times 10^{-8} \, \rm  ph \, cm^{-2} \, s^{-1}$. 
The photon index variability is dominated by errors, since the variance $\sigma^2$ is smaller than the mean square uncertainty $\delta^2$ (see Sect.\ 4).
In Fig.\ \ref{multi} we can see the corresponding $\gamma$-ray light curve (red points).
We also plotted a monthly-binned light curve (blue points) that includes many upper limits (cyan points) because of the source faintness. Finally, during the 2013 outburst we performed weekly bins when there was a good count number to detail the flux variations (green points). 

\section{Correction for Galactic and DLA extinction}
\label{asso}

In the previous sections we presented light curves of OJ~248 as observed magnitudes.
But the near-IR, optical, and UV radiation from the source suffers absorption by both the Galaxy dust and the dust contained in the intervening DLA system at $z=0.525$ mentioned in the Introduction. This is a problem similar to that met when analyzing data from another well-known blazar, AO 0235+16 \citep{rai05}.

We estimated the Galactic reddening in the $UBVRIJHK$ optical and near-IR bands by using the \citet{car89} extinction law, with $R_V=A_V/E(B-V)=3.1$, which is the standard value for the diffuse interstellar medium.
The results are reported in Table \ref{ext} (Column 2).
An estimate of the extinction due to the DLA system can be calculated starting from the hydrogen column density $N_H=(2.0 \pm 0.2) \times 10^{20} \rm \, cm^{-2}$ obtained by \citet{rao00} and adopting a gas-to-dust ratio equal to the average value in the Milky Way: $N_H=4.93 \times 10^{21} {\rm \, cm^{-2} \, mag^{-1}} \times E(B-V)$ \citep{dip94}. This yields $E(B-V)=0.04$, and assuming again $R_V=3.1$, we obtain $A_V=0.124$ at $z=0.525$.
Then, by applying the same \citet{car89} law properly blueshifted, we get 
the values reported in Table \ref{ext} (Column 3).
We note that the DLA system is a more important absorber than the Galaxy.
The assumption that the DLA system has the same absorbing characteristics of the Milky Way is justified by its being a spiral galaxy \citep{rao03}.
The values of the total extinction that we must apply to correct our data for absorption in both the Galaxy and the DLA system are given in Column 4.

In the case of the {\it Swift}-UVOT bands, because of the asymmetric shape of the filter responses and of the bumped shape of the extinction law in the UV, we calculated the absorption in the various bands by integrating the \citet{car89} law with the filter effective areas \citep[see e.g.][]{rai10}. The results are shown in Table \ref{ext}.

\begin{table}
 \centering
  \caption{Extinction [mag] in the various Bessel \citep{bes98} and {\it Swift}-UVOT bands toward OJ~248. 
Both the Galactic absorption and that by the DLA system at $z=0.525$ are given.
The value of the total extinction suffered by the source radiation is the sum of the two.}
   \label{ext}
  \begin{tabular}{@{}cccc@{}}
  \hline
Band & $A_\lambda$(Gal) & $A_\lambda$(DLA) & Total\\
 \hline
\multicolumn{4}{c}{{\it Swift}-UVOT bands}\\
$w2$ & 0.249 & 0.397 & 0.646\\
$m2$ & 0.261 & 0.343 & 0.604\\
$w1$ & 0.211 & 0.331 & 0.542\\
$u$ & 0.150 & 0.349 & 0.499\\
$b$ & 0.125 & 0.240 & 0.365\\
$v$ & 0.095 & 0.196 & 0.291\\
\hline
\multicolumn{4}{c}{Bessel bands}\\
$U$ & 0.142 & 0.316 & 0.458\\
$B$ & 0.122 & 0.235 & 0.357\\
$V$ & 0.093 & 0.195 & 0.288\\
$R$ & 0.077 & 0.173 & 0.250\\
$I$ & 0.055 & 0.131 & 0.186\\
$J$ & 0.027 & 0.074 & 0.101\\
$H$ & 0.017 & 0.045 & 0.062\\
$K$ & 0.011 & 0.028 & 0.039\\ 
\hline
\end{tabular}
\end{table}

\section{Cross-correlation between variability at high and low energies}

\begin{figure}
\includegraphics[width=8.5cm]{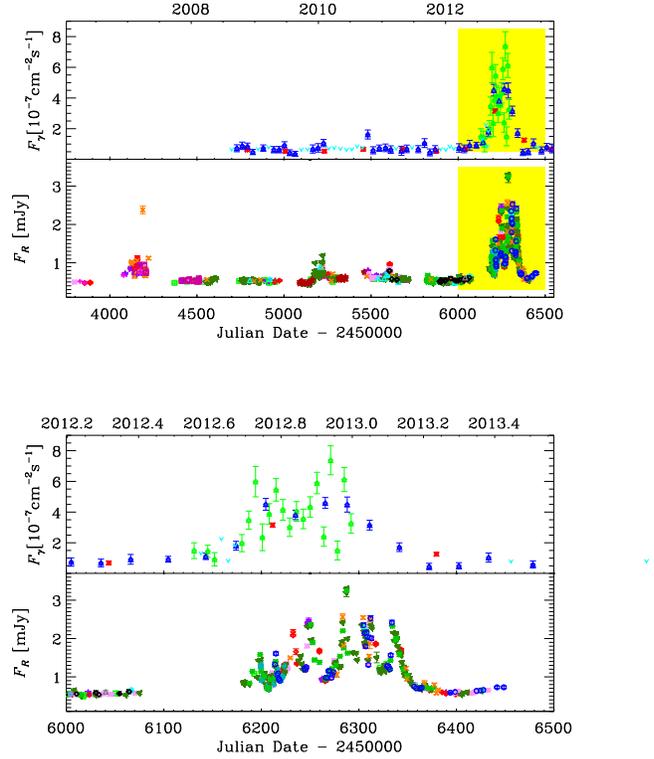}
  \caption{Light curves in the $\gamma$-ray and optical ($R$) bands over the whole period (top), and during the 2012--2013 outburst period indicated by the yellow stripe (bottom).}
\label{zoom}
\end{figure}

In order to better investigate the relationship between the source behaviour in $\gamma$-rays and that in the optical band, we show in Fig.\ \ref{zoom} the corresponding whole light curves as well as an enlargement of the 2012--2013 outburst period, where the correlation is easier to study. The start of the $\gamma$-ray outburst is not covered in the optical band because of the solar conjunction, but the first point after the seasonal gap is about 0.3 mag brighter than before, suggesting that the outburst has already begun.
In contrast, the optical light curve is very well sampled in the outburst decline phase, where the $\gamma$-ray curve has a worst time resolution because of the low flux.
It seems that the $\gamma$ and optical fluxes may well have risen and declined together, but that the period of major $\gamma$ activity preceded the phase of strongest optical activity.

\begin{figure}
\includegraphics[width=8.5cm]{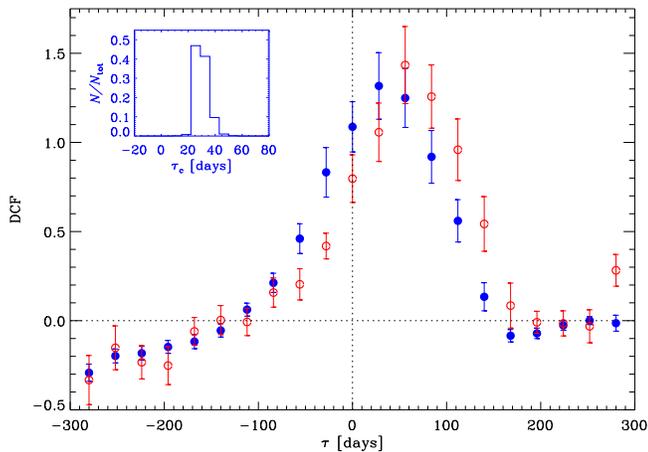}
  \caption{Discrete correlation functions between the $\gamma$-ray fluxes and the $R$-band de-absorbed flux densities (blue filled circles) and between the $\gamma$-ray fluxes and the X-ray flux densities at 1 keV (red empty circles). The inset shows the results of Monte Carlo simulations of the $\gamma$-optical correlation (see text for details).}
\label{cross}
\end{figure}

To get a quantitative estimate of this time shift, we calculate the DCF (see Sect.\ \ref{radiomm}) between the $\gamma$-ray fluxes and the $R$-band flux densities corrected for the total absorption reported in Table \ref{ext}\footnote{We considered the weekly $\gamma$ fluxes during the outburst period, and the monthly fluxes before and after that (see Fig.\ \ref{zoom}), while we binned the optical data in seven-day bins.}. 
The DCF is displayed in Fig.\ \ref{cross}. It shows a peak at a lag of $\tau_{\rm p} = 28$ days with $\rm DCF_{p} = 1.3$ that indicates strong correlation with the optical variations following the $\gamma$-ray ones after four weeks. The delay is 29 days if we take the centroid instead of the peak.
The figure inset displays the result of 2000 Monte Carlo simulations according to the ``flux randomization-random subset selection" method \citep{pet98,rai03}.
From these simulations it is possible to estimate the uncertainty on the delay.
We obtained that 88\% of the realizations led to a centroid value between 22 and 36 days.
Hence, we infer that  the optical flux variations follow the $\gamma$ flux changes with a delay of  $29 \pm 7 \rm \, days$. 

However, if the $\gamma$ emission is due to inverse-Compton scattering of soft photons off the same electrons producing the optical radiation, then its variations are expected to be simultaneous or delayed with respect to those characterising the optical radiation, as resulting from
modelling non-thermal flares with shocks in a jet \citep[e.g.][]{sikora01,sokolov04,sokolov05}. This was observed in several blazars, in particular in the FSRQs 4C~38.41 \citep{rai12}, 3C~345 \citep{schinzel12}, and 3C~454.3 \citep{bon09,ver10,rai11}.
In contrast, $\gamma$ variations leading the optical ones were observed e.g.\ in the FSRQs PKS~1510-09 by \citet{abdo10_1510} and \citet{dam11}, and 3C~279 by \citet{hay12}. The latter authors found a lag of about 10 days and explained it by assuming that the energy density of the external seed photons for the inverse-Compton process decreases faster along the jet than the energy density of the magnetic field causing the synchrotron optical emission \citep[see also][]{janiak12}. Analogous interpretations may also hold for OJ~248.

Alternatively, the complex optical/$\gamma$-ray correlation may be explained by considering the effects of turbulence in the jet \citep{mar14}.
The fluctuating magnitude and direction of a turbulent magnetic field affects mostly the synchrotron radiation, and therefore adds a component to the optical variability that is not present in the gamma-ray light curve.

Figure \ref{cross} also displays the DCF between the $\gamma$-ray fluxes and the X-ray flux densities at 1 keV, suggesting a delay of the X-ray variations of about 2 months. We finally investigated the $\gamma$-mm correlation, finding a strong signal and a $\sim 70$ days delay of the mm variations.

\begin{figure}
\includegraphics[width=8.5cm]{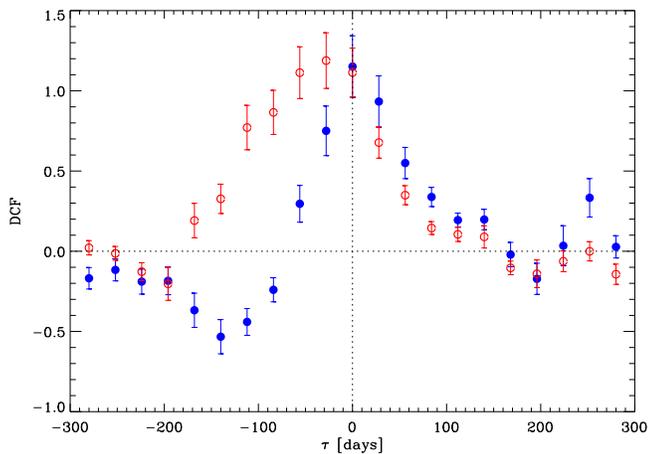}
  \caption{Discrete correlation functions between the X-ray and 230 GHz flux densities (blue filled circles) and between the X-ray and $R$-band de-absorbed flux densities (red empty circles).}
\label{xmm}
\end{figure}

Another interesting correlation holds between the X-ray and mm flux densities. Indeed, beside the 2012--2013 outburst, the X-ray light curve also shows a peak during the maximum of the mm light curve at the end of 2010 (see Fig.\ \ref{multi}). The DCF in Fig.\ \ref{xmm} displays a strong correlation with no time lag, which suggests that the X-ray and mm radiation are produced in the same region, with the X-ray emission likely due to inverse-Compton on the mm photons.
A correlation between the X-ray and mm variability has already been found, e.g.\ in BL Lacertae \citep{rai13}.

In Fig.\ \ref{xmm} we also show the DCF between the X-ray and optical flux densities, indicating that the optical variations precede those in the X-rays by about one month.

Cross-correlations of the $\gamma$-ray, X-ray, and $R$-band data with the 37 GHz data only led to weak and somewhat confused signals. This is due to the different behaviour of the corresponding light curves, in particular to the dominance of the 2010--2011 outburst with respect to the 2013 one at 37 GHz.

\section{Broad-band SED}

\begin{figure*}
\centering
\includegraphics[width=14cm]{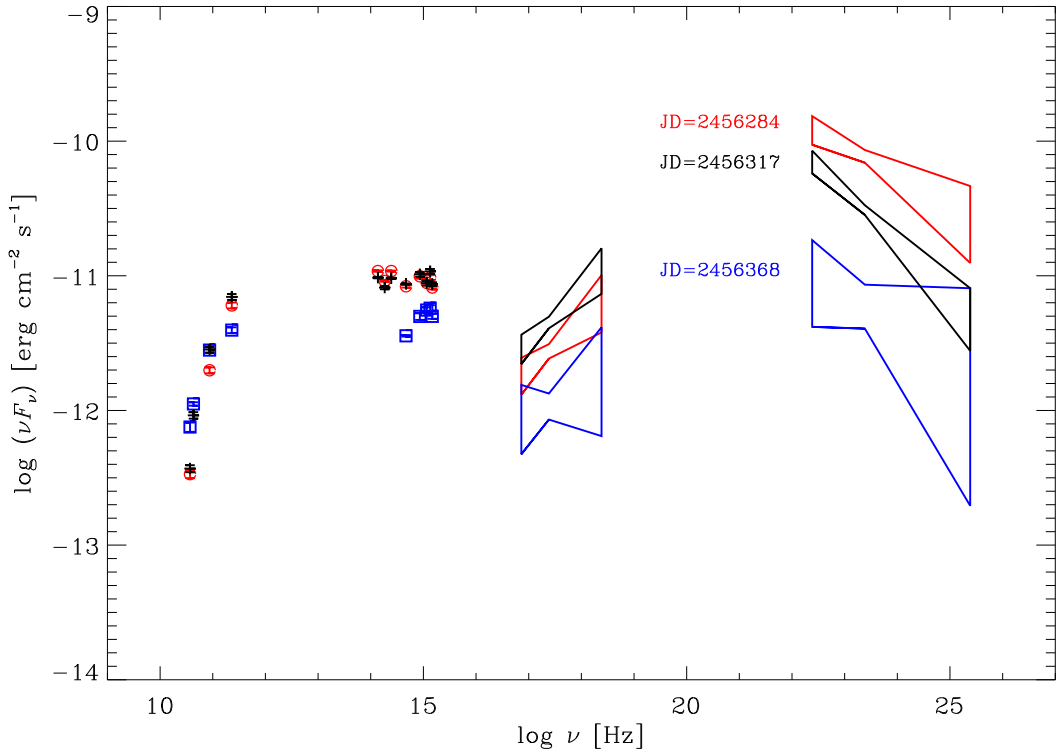}
  \caption{Spectral energy distributions of OJ~248 from the radio to the $\gamma$-ray frequencies during three epochs characterised by different brightness states.}
\label{sed}
\end{figure*}

Three broad-band SEDs of OJ~248 are plotted in Fig.\ \ref{sed}. 
They correspond to the peak of the $\gamma$-ray emission ($\rm JD=2456284$), to the peak of the X-ray emission ($\rm JD=2456317$), and to a faint post-outburst epoch ($\rm JD=2456368$).
The SEDs are built with simultaneous near-IR, optical, UV, X-ray, and $\gamma$-ray data.
Because of the smoother radio variability, we gave a tolerance of a few days to the radio data.

Emission in the optical--UV receives an important contribution from the accretion disc radiation, whose signature is more evident in the faint, post-outburst SED. The concave shape of the near-IR spectrum is due to the intersection between the disc contribution and the non-thermal jet emission. These two components have been modelled by \citet{rai14}, who found that the OJ~248 disc is more luminous than a typical type 1 QSO disc.

Notice that the faintest SED at high energies has also the lowest radio flux at 230 GHz, but it exceeds the fluxes of the other SEDs at longer radio wavelengths.

\section{Summary and Conclusions}

In this paper we have presented the results of a huge multiwavelength observing effort led by the GASP-WEBT Collaboration on the blazar OJ~248. Data were collected starting from 2006 and up to 2013, including two optical-NIR outbursts in 2006--2007 and 2012--2013 and two major radio outbursts in 2010--2011 and 2012--2013. The 2012--2013 outburst was also detected at high energies by the {\it Swift} and {\it Fermi} satellites. The correlation between the optical and radio outbursts is clear in 2012--2013, while the optical counterpart of the 2010--2011 radio outburst is difficult to identify. Something likely changed in the source in the period between the two outbursts, one possibility being a slightly better alignment of the optical emitting region with the line of sight with the consequent increase of the Doppler beaming.
A strong correlation between the flux variations at $\gamma$-rays and those in the optical band is found, but with the optical variations delayed by about one month, which is a peculiar behaviour already found in other blazars. Strong correlation with no time delay has also been found between the X-ray and millimetre flux changes, supporting a common emission region in the jet.

We have analysed the polarimetric behaviour of the source. The fraction of polarised flux remained low for most of time but during the 2012--2013 outburst, when $P$ reached $\sim 19\%$. Wide rotations of the linear polarisation vector can reliably be detected only during the outburst and they occur in both directions, suggesting a complex behaviour of the magnetic field in the jet possibly due to turbulence, and/or a complex jet structure involving spiral paths. 

Optical spectra show Mg~II lines both in absorption and emission. The absorption line is due to an intervening system at z=0.525 whose reddening effects on the NIR, optical, and UV source emission have been estimated and taken into account. The presence of the intervening system must be considered also when analysing the X-ray radiation. As for the Mg~II emission line from the source BLR, we estimated a mean velocity of $(2053 \pm 310) \rm \, km \, s^{-1}$. The line flux is essentially stable around a mean value of $(6.2 \pm 0.5) \times 10^{-15} \rm \, erg \, cm ^{-2} \, s^{-1}$ also during the 2012--2013 outburst and after, confirming that the jet emission did not affect the BLR, even when considering up to a few months of possible time delay.

\section*{Acknowledgments}
The data collected by the GASP-WEBT collaboration are stored in the GASP-WEBT archive; for questions regarding their availability, please contact the WEBT President Massimo Villata ({\tt villata@oato.inaf.it}).
We thank the referee, Vasiliki Pavlidou, for useful comments and suggestions.
This article is partly based on observations made with the telescopes IAC80 and TCS operated by the Instituto de Astrofisica de Canarias in the Spanish Observatorio del Teide on the island of Tenerife. Most of the observations were taken under the rutinary observation program. The IAC team acknowledges the support from the group of support astronomers and telescope operators of the Observatorio del Teide. 
St.Petersburg University team acknowledges support from Russian RFBR grant 15-02-00949 and St.Petersburg University research grant 6.38.335.2015.
AZT-24 observations are made within an agreement between  Pulkovo, Rome and Teramo observatories.
The Abastumani team acknowledges financial support of the project
FR/638/6-320/12 by the Shota Rustaveli National Science Foundation under
contract 31/77.
This paper is partly based on observations carried out at the German-Spanish Calar Alto Observatory, which is jointly operated by the MPIA and the IAA-CSIC.
Acquisition and reduction of the MAPCAT and IRAM 30 m data is supported in part by MINECO (Spain) grant and AYA2010-14844, and by CEIC (Andaluc\'{i}a) grant P09-FQM-4784.
This paper is partly based on observations carried
out at the IRAM 30 m Telescope, which is supported by INSU/CNRS (France), MPG (Germany), and IGN (Spain).
The Submillimeter Array is a joint project between the Smithsonian 
Astrophysical Observatory and the Academia Sinica Institute of Astronomy 
and Astrophysics and is funded by the Smithsonian Institution and the 
Academia Sinica. 
The Steward Observatory spectropolarimetric monitoring project is supported by the Fermi Guest Investigator grants NNX08AW56G and NNX09AU10G.
The research at Boston University (BU) was funded in part by NASA Fermi Guest Investigator grants NNX11AQ03G and NNX13AO99G.
The PRISM camera at Lowell Observatory was developed by K.\ Janes et al. at BU and Lowell Observatory, with funding from the NSF, BU, and Lowell Observatory.
The Liverpool Telescope is operated on the island of La Palma by Liverpool John Moores University in the
Spanish Observatorio del Roque de los Muchachos of the Instituto de Astrofisica de Canarias, with funding
from the UK Science and Technology Facilities Council.
The Mets\"ahovi team acknowledges the support from the Academy of Finland
to our observing projects (numbers 212656, 210338, 121148, and others).
Partly based on observations with the Medicina and Noto telescopes
operated by INAF - Istituto di Radioastronomia.
E.~ B.\ acknowledge financial support from UNAM-DGAPA-PAPIIT
through grant IN111514.
GD and OV did the observations/investigations in line with Projects No. 176011 (Dynamics and Kinematics of Celestial Bodies and Systems), 176021 (Visible and Invisible Matter in Nearby Galaxies: Theory and Observations) and 176004 (Stellar Physics) which are supported by the Ministry of Education, Science and Technological Development of the Republic of Serbia.
The Torino team acknowledges financial support by INAF through contract PRIN-INAF 2011.

\newpage
\noindent{\large \bf AFFILIATIONS}

\medskip\noindent
{\it
$^{ 1}$INAF, Osservatorio Astrofisico di Torino, via Osservatorio 20, Pino Torinese, Italy    \\                                                 
$^{ 2}$Instituto de Astrofisica de Canarias (IAC), La Laguna, Tenerife, Spain                              \\
$^{ 3}$Departamento de Astrofisica, Universidad de La Laguna, La Laguna, Tenerife, Spain                   \\
$^{ 4}$Dip. Fisica e Astronomia, Universit\`a degli Studi di Bologna, Via Ranzani 1, I-40127, Bologna, Italy\\
$^{ 5}$INAF - Istituto di Radioastronomia, Via Gobetti 101, I-40129 Bologna, Italy                         \\
$^{ 6}$Steward Observatory, University of Arizona, Tucson, AZ 85721, USA                                   \\
$^{ 7}$Astron.\ Inst., St.-Petersburg State Univ., Russia                                                  \\
$^{ 8}$Pulkovo Observatory, St.-Petersburg, Russia                                                         \\
$^{ 9}$Isaac Newton Institute of Chile, St.-Petersburg Branch                                              \\
$^{10}$Instituto de Astrof\'{\i}sica de Andaluc\'{\i}a (CSIC), Apartado 3004, E-18080 Granada, Spain       \\
$^{11}$Current Address: Joint Institute for VLBI in Europe, Postbus 2, NL-7990 AA Dwingeloo, the Netherlands\\
$^{12}$Max-Planck-Institut f\"ur Radioastronomie, Auf dem H\"ugel, 69, D-53121, Bonn, Germany              \\                                          
$^{13}$Institute of Astronomy, Bulgarian Academy of Sciences, 1784 Sofia, Bulgaria                         \\
$^{14}$Instituto de Astronom\'ia, Universidad Nacional Aut\'onoma de M\'exico, Apdo. Postal 70-264, M\'exico\\
$^{15}$Physics  Department, University of Crete, Heraklion, Greece                                         \\
$^{16}$Dept.\ of Astronomy, Faculty of Physics, Sofia University, Bulgaria                                 \\
$^{17}$INAF, Osservatorio Astrofisico di Catania, Italy                                                    \\
$^{18}$EPT Observatories, Tijarafe, La Palma, Spain,                                                       \\
$^{19}$INAF, TNG Fundaci\'on Galileo Galilei, La Palma, Spain                                              \\
$^{20}$Graduate Inst.\ of Astronomy, National Central University, 300 Zhongda Road, 32001 Zhongli, Taiwan  \\
$^{21}$Astronomical Observatory, Volgina 7, 11060 Belgrade, Serbia                                         \\
$^{22}$INAF, Osservatorio Astronomico di Roma, Italy                                                       \\
$^{23}$Maidanak Observatory of the Ulugh Beg Astronomical Institute, Uzbekistan                            \\
$^{24}$Harvard-Smithsonian Center for Astrophysics, Cambridge MA, 02138 USA                                \\
$^{25}$Instituto de Astronom\'ia, Universidad Nacional Aut\'onoma de M\'exico, Ensenada, B.C., 22800, M\'exico\\
$^{26}$Institute for Astrophysical Research, Boston University, 725 Commonwealth Avenue, Boston, MA 02215, USA\\
$^{27}$Abastumani Observatory, Mt. Kanobili, Abastumani, Georgia \\
$^{28}$ZAH, Landessternwarte Heidelberg, Königstuhl 12, 69117 Heidelberg, Germany  \\
$^{29}$Engelhardt Astronomical Observatory, Kazan Federal University, Tatarstan, Russia \\
$^{30}$Center for Astrophysics, Guangzhou University, Guangzhou 510006, China \\
$^{31}$Aalto University Mets\"ahovi Radio Observatory, Mets\"ahovintie 114, 02540 Kylm\"al\"a, Finland     \\
$^{32}$Aalto University Department of Radio Science and Engineering, P.O. BOX 13000, FI-00076 AALTO, Finland\\
$^{33}$Dept.\ of Physics $\&$ Astronomy, University of Canterbury, Christchurch, New Zealand                  \\
$^{34}$Dept.\ of Physics and Astronomy, Univ.\ of Southampton, Southampton, SO17 1BJ, United Kingdom       \\
$^{35}$Armagh Observatory, College Hill, BT61 9DG, Armagh, N. Ireland     \\                                 
$^{36}$Astrophysikalisches Institut Potsdam, An der Sternwarte 16, D-14482 Potsdam, Germany         \\      
$^{37}$Lowell Observatory, Flagstaff, AZ 86001, USA    \\                                                  
$^{38}$Space Science Institute, Boulder, CO 80301, USA \\
}

\bsp

\label{lastpage}

\end{document}